\documentclass[conference]{IEEEtran}

\makeatletter
\def\ps@headings{%
\def\@oddhead{\mbox{}\scriptsize\rightmark \hfil \thepage}%
\def\@evenhead{\scriptsize\thepage \hfil \leftmark\mbox{}}%
\def\@oddfoot{}%
\def\@evenfoot{}}
\makeatother
\usepackage{algorithm}
\usepackage{algorithmicx}
\usepackage{algpseudocode}
\usepackage{amsmath}
\usepackage{graphicx}
\usepackage{subfigure}
\usepackage{times}
\usepackage[hyphens]{url}
\usepackage{color}
\usepackage{multirow}
\usepackage{textcomp}
\usepackage{threeparttable}
\usepackage{supertabular}
\usepackage{amssymb}
\usepackage{amsfonts}
\usepackage{booktabs}
\usepackage{epstopdf}
\usepackage{xspace}
\usepackage{enumitem}
\usepackage{cite}
\usepackage{fancyhdr}

\newcommand{\FakeSwarm}{{\sc FakeSwarm}\xspace}

\newcommand{\ISOT}{{\sc ISOT Fake News}\xspace}

\renewcommand{\paragraph}[1]{\smallskip\noindent {\bf #1}}

\begin{document}

%----------------------------------------------------------------------------
\title{FakeSwarm: Improving Fake News Detection with Swarming Characteristics}

\author{\IEEEauthorblockN{1\textsuperscript{st} Jun Wu$^*$}
\IEEEauthorblockA{
%\textit{College of Computing} \\
\textit{Georgia Institute of Technology}\\
Atlanta, United States \\
jwu772@gatech.edu}
\and
\IEEEauthorblockN{2\textsuperscript{nd} Xuesong Ye}
\IEEEauthorblockA{
%\textit{Department of Engineering and Technology} \\
\textit{Trine University}\\
Phoenix, United States \\
xye221@my.trine.edu}
}

\maketitle

\thispagestyle{fancy}
\fancyhead{}
\lhead{}
\lfoot{979-8-3503-3698-6/23/\$31.00 ©2023 IEEE \hfill}
\cfoot{}
\rfoot{}

\begin{abstract}
The proliferation of fake news poses a serious threat to society, as it can misinform and manipulate the public, erode trust in institutions, and undermine democratic processes. To address this issue, we present FakeSwarm, a fake news identification system that leverages the swarming characteristics of fake news. To extract the swarm behavior, we propose a novel concept of fake news swarming characteristics and design three types of swarm features, including principal component analysis, metric representation, and position encoding. We evaluate our system on a public dataset and demonstrate the effectiveness of incorporating swarm features in fake news identification, achieving an f1-score and accuracy of over 97\% by combining all three types of swarm features. Furthermore, we design an online learning pipeline based on the hypothesis of the temporal distribution pattern of fake news emergence, validated on a topic with early emerging fake news and a shortage of text samples, showing that swarm features can significantly improve recall rates in such cases. Our work provides a new perspective and approach to fake news detection and highlights the importance of considering swarming characteristics in detecting fake news.
\end{abstract}


\section{Introduction}

\label{sec:intro}
\subsection{Background and Motivation}
Fake news poses a significant risk to users of social media platforms, as it can cause psychological and health-related issues, increase polarization, and erode public trust. The vast amounts of false information available on these platforms, coupled with their ability to spread rapidly and without restriction, can lead to widespread panic and anxiety among users. Studies have shown that exposure to fake news can cause stress, depression, and other mental health problems. Furthermore, the use of text generation technologies and DeepFake techniques, have made it increasingly challenging to differentiate between authentic and fake news. As such, there is a growing need to develop machine learning algorithms that can detect and combat fake news. Recent research has shown that natural language processing techniques, social network analysis, and deep learning can be used to identify fake news with high accuracy. However, much work remains to be done in this field.

\subsection{Fake News Detection}
\label{subsec:related}
Several surveys and overviews have been conducted to provide a comprehensive understanding of the fake news phenomenon, its characteristics, detection methodologies, and future research opportunities.
Zhou \emph{et al.} \cite{zhou2020survey} provide a survey covering fundamental theories, detection methods including machine learning and NLP techniques, and future research opportunities for fake news, emphasizing the impact of fake news on users and the importance of developing innovative detection methods. Zhang \emph{et al.} \cite{zhang2020overview} offer a comprehensive analysis of online fake news, discussing its characteristics and detection methodologies, and fostering academic discussion around potential solutions, providing insights into the evolving landscape of fake news and its consequences. Shu \emph{et al.} \cite{shu2019beyond} emphasize the importance of social context in fake news detection by examining the impact on users and exploring a range of detection methods that consider user behavior and interactions, highlighting the role of social dynamics in identifying deceptive content.

The existing fake news detection research can be roughly divided into two categories: \emph{user perspective} and \emph{content based}. The first category focuses on user behavior, network structure, and user profile features to model the correlation between user activities and the spread of fake news. The second category leverages various types of data, such as text, images, and metadata, to provide a more accurate and holistic understanding of potentially deceptive content. 
 
\subsubsection{User Perspective} 
User perspective features, such as user behavior and network structure, play an essential role in detecting fake news on social media platforms. They consider the user's role in spreading fake news, user interactions and relationships within social networks, and user profile features, including demographics, interests, and history. By analyzing these features, researchers can gain a deeper understanding of user behavior and more accurately identify potential fake news content.
Shu \emph{et al.} \cite{shu2017fake} examine user perspective features, such as user behavior and network structure, to detect fake news on social media using data mining techniques and various recognition models, considering the user's role in spreading fake news. Monti \emph{et al.} \cite{monti2019fake} utilize geometric deep learning to analyze user social graphs for fake news detection, emphasizing user interactions and relationships within social networks, enabling a more accurate assessment of the likelihood of fake news based on users' involvement. Shu \emph{et al.} \cite{shu2018understanding} develop a fake news detection model focusing on user profile features, including demographics, interests, and history, which allows for a deeper understanding of user behavior and more accurate identification of potential fake news content. Ruchansky \emph{et al.} \cite{ruchansky2017csi} propose a hybrid deep learning model that analyzes temporal patterns in user activities to effectively detect fake news on social media platforms, considering the timing of user interactions and engagements for fake news identification. Sahoo \emph{et al.} \cite{sahoo2021multiple} create a robust fake news detection model on social networks by combining deep learning with user involvement features, such as connections and interactions with others, emphasizing the importance of understanding user dynamics and relationships.

\subsubsection{Content Based} 

In recent years, researchers leveraged text, image and multimodal data to detect fake news. These methods combine various types of data, such as text, images, and metadata, to provide a more accurate and holistic understanding of potentially deceptive content.
Khattar \emph{et al.} \cite{khattar2019mvae} introduce a multi-modal variational autoencoder (MVAE) model that fuses textual and visual information for fake news detection, offering a comprehensive approach to understanding and detecting misleading content using both text and image data. Singhal \emph{et al.} \cite{singhal2019spotfake} propose a multi-modal framework utilizing text, images, and metadata in combination with deep learning models for effective fake news detection, leveraging various types of data to provide a more accurate and holistic understanding of potentially deceptive content. Granik \emph{et al.} \cite{granik2017fake} employ a Naive Bayes classifier for fake news detection based on textual features and probabilistic reasoning, considering the likelihood of text being fake news and offering a statistical method for identification. Kaliyar \emph{et al.} \cite{kaliyar2020fndnet} present a deep convolutional neural network (CNN) model for fake news detection that analyzes text with high accuracy, leveraging the strengths of CNNs in text analysis and feature extraction to detect deceptive content. Karimi \emph{et al.} \cite{karimi2018multi} develop a model using advanced NLP and machine learning techniques to detect fake news by combining multiple sources and classes of information, recognizing the importance of considering various aspects of content and user behavior. Wang \emph{et al.} \cite{wang2018eann} propose an event adversarial neural network that leverages multi-modal data (text, images, and metadata) to improve fake news detection performance, utilizing multiple data types to offer a comprehensive understanding of content for more accurate detection. Kaliyar \emph{et al.} \cite{kaliyar2021fakebert} utilize a BERT-based deep learning model for effective detection and classification of fake news in social media content, leveraging the power of BERT for a highly accurate and context-aware fake news detection method.

\subsection{Attack and Defense in Other Layers}

Most research proves that sophisticated attacking methods exist in layers beyond the fake content itself. These methods involve the utilization of \emph{social bots} to breach social media platforms and the use of \emph{adversarial learning} and \emph{multimedia data generation} techniques to confuse detection models.

\subsubsection{Social bots}

Online social applications are vital platforms for people to share and read news. However, bot accounts are flooding almost every popular social media application, and many research works have started to reveal their harmful activities, such as spreading fake news, rumors, and misinformation. Emilio \emph{et al.} \cite{ferrara2016rise} revealed the rising process of social bots, which are software agents developed by the black industry to mimic humans. Stefano \emph{et al.} \cite{cresci2017paradigm} investigated account performance discrimination between genuine accounts and bot accounts on Twitter. Giovanni \emph{et al.} \cite{santia2019detecting} explored the negative influence of social bots spreading misinformation on Facebook.

Detecting and deleting fake news from the platform is a direct defense strategy, but discovering and banning bot accounts can solve the problem at its root. Maryam \emph{et al.} \cite{heidari2022online} built account embeddings to transform metadata, such as age, gender, and personality, into a feature space for bot classification. Shangbin \emph{et al.} \cite{feng2022heterogeneity} investigated heterogeneity structures in the account relation graph, spreading information among nodes, and discriminating between genuine users and social bots. BotShape \cite{wu2023botshape} built a novel and accurate detection system based on the disparity of account behavioral patterns between bots and real humans. BotTriNet \cite{wu2023bottrinet} is a content model for bot detection that first applies metric learning to increase the distance between bots and normal users in the embedding space, resulting in significant accuracy improvement for content-less bot categories. BotMoE \cite{feng2022heterogeneity} is a Twitter bot detection system that uses multiple modalities, such as textual content and network topology, to improve detection accuracy.

\subsubsection{Model Security}
We consider two aspects of model security: (i) the robustness of the detection model against evasion attempts and (ii) the speed of detection to prevent the spread of fake news. Model attacks have dual purposes: they provide a way to bypass model identification and also offer opportunities to enhance model robustness. The black industry leverages multimedia data generation and adversarial learning technologies to evade detection.

For textual data generation, Rik \emph{et al.} \cite{koncel2019text} used a Transformer to generate text from a knowledge graph. Zhuoyi \emph{et al.} \cite{wang2021contextual} designed a novel rephrase detection system based on contextual content in dialogue scenes. Image generation and synthesis techniques are widely utilized. Tao \emph{et al.} \cite{zhang2022deepfake} provided a systematic introduction to DeepFake generation and detection. LiveBugger \cite{li2022seeing} systematically studied the security of facial liveness verification systems and improved the attack success rate by up to 70\%. Qingzhao \emph{et al.} \cite{zhang2022adversarial} investigated adversarial examples to exploit prediction errors in vehicle trajectories. Jiachen \emph{et al.} \cite{sun2020towards} proposed a novel black-box adversarial sensor attack targeting the security of autonomous driving perception models. To defend against adversarial examples, Changjiang \emph{et al.} \cite{li2019det} designed a transferability-based approach for both black and gray box attacks on deep neural networks.

Real-time detection of fake news is crucial because online users can immediately consume them once they are published. Knowledge distillation is the process of transferring knowledge from a complex model to a simpler model, especially for deep models. Jianping \emph{et al.} \cite{gou2021knowledge} provide a comprehensive survey of knowledge distillation techniques. Yuke \emph{et al.} \cite{zhang2023c2pi} proposed a distillation-based inverse-network attack via partitioning the neural network model. Souvik \emph{et al.} \cite{kundu2023making} designed a model compression method by removing ReLU layers and merging them with preceding layers.

\subsection{Contributions} 
To summarize, this paper makes the following contributions: 

\begin{itemize} 
%设计了FakeSwarm的fake news识别系统，创新地提出了fake news团伙特征的概念，并设计了三类通用特征和提取方法，分别是主要成分、metric表示、和位置编码。

%在公开数据集上，分别验证了三类团伙特征，对原始文本分类任务的精度提升效果，并综合所有团伙特征后，取得了97%的f1score和accuracy。验证了团伙特性对识别精度的贡献。

%基于对fake news的涌现的时间分布规律的基本假设，设计了online learning的pipeline, 并验证了某个topic的fake news早期出现，且文本样本不足的时候，团伙特征对提高召回率有很大作用。

\item We designed FakeSwarm, a fake news identification system that incorporates an innovative concept of fake news swarming characteristics. The system includes three types of generic features and extraction methods, namely principal component analysis, metric representation, and position encoding.

\item We evaluated the three types of swarm features on a public dataset and analyzed their contributions to the classification accuracy of the original text. By combining all three types of swarm features, our system achieved an f1-score and accuracy of over 97\%, demonstrating the effectiveness of incorporating swarm features in fake news identification.

\item Based on the hypothesis of the temporal distribution pattern of fake news emergence, we designed an online learning pipeline. We validated the pipeline on a topic with early emerging fake news and a shortage of text samples, demonstrating that swarm features can significantly improve recall rates in such cases.

\end{itemize}

\section{Dataset}
\label{sec:dataset}
%https://onlineacademiccommunity.uvic.ca/isot/wp-content/uploads/sites/7295/2023/02/ISOT_Fake_News_Dataset_ReadMe.pdf
Ahmed {\em et al.} \cite{ahmed2017detection} and \cite{ahmed2018detecting} designed text-based fake news detection systems based on a real-world fake news dataset they collected called \ISOT. The dataset includes news articles from 2015 to 2018, with 23,481 fake news and 21,417 real news articles. The fake news articles were collected from different sources flagged by fact-checking organizations like Politifact and Wikipedia. In contrast, truthful articles were obtained by crawling articles from Reuters.com.

Each news article in the dataset is provided with the title, body text, subject, published date, and a label that distinguishes between fake and real news. The fake news articles cover multiple subjects, including Left News, Politics, Middle-East, News, Government News, and US News, while the real news articles are categorized into Politics News and World News. Categories of fake news and real news are mutually inclusive, meaning that subjects are not useful information for distinguishing between fake and real news.

We observed that the sources of some texts were correlated with the labels. For example, body texts with a beginning of Reuters were always real news. To prevent source information from leaking the labels, we removed the source information from each body text in \ISOT, making the dataset as objective and unbiased as possible.

\section{Design}
We present \FakeSwarm, a fake news detection system leveraging the swarming characteristics of a few news to improve the detection accuracy. It takes the body texts of news as input data and predicts whether the news is fake or not.

%解释什么是swarming？ 
The approach leverages the concept of "swarming", which refers to the tendency of fake news to spread quickly and widely across online platforms. A swarm is a group of fake news with similar topics that occur in a concentrated period of time.

%大概解释用什么方法，以及理由
To enhance the swarming attributes in the original text embeddings learned from news body texts, \FakeSwarm designed three approaches: (i) principal embedding produced by principal components analysis (PCA); (ii) metric embedding produced by contrastive learning; (iii) position embedding produced by a clustering algorithm. Principal embedding is the projection of raw text embedding with the most important and reduced dimensionality, which actively pushes data points in a swarm closer. Metric learning is a representation of raw text embedding learned by unsupervised contrastive learning loss, which actively increases the distance between fake news and real news and also reduces the distance between fake news and swarms. Position embedding stands for the distances between news to all clustering centers of text embeddings, respectively. If the news is close to an arbitrary clustering center, the probability of belonging to a swarm will increase.

We will first introduce how to transform raw news body texts into text embeddings. And then, we explain how these swarming characteristic embedding mining approaches and their implementation.

\begin{figure}[htb]
\centering
\includegraphics[width=\linewidth]{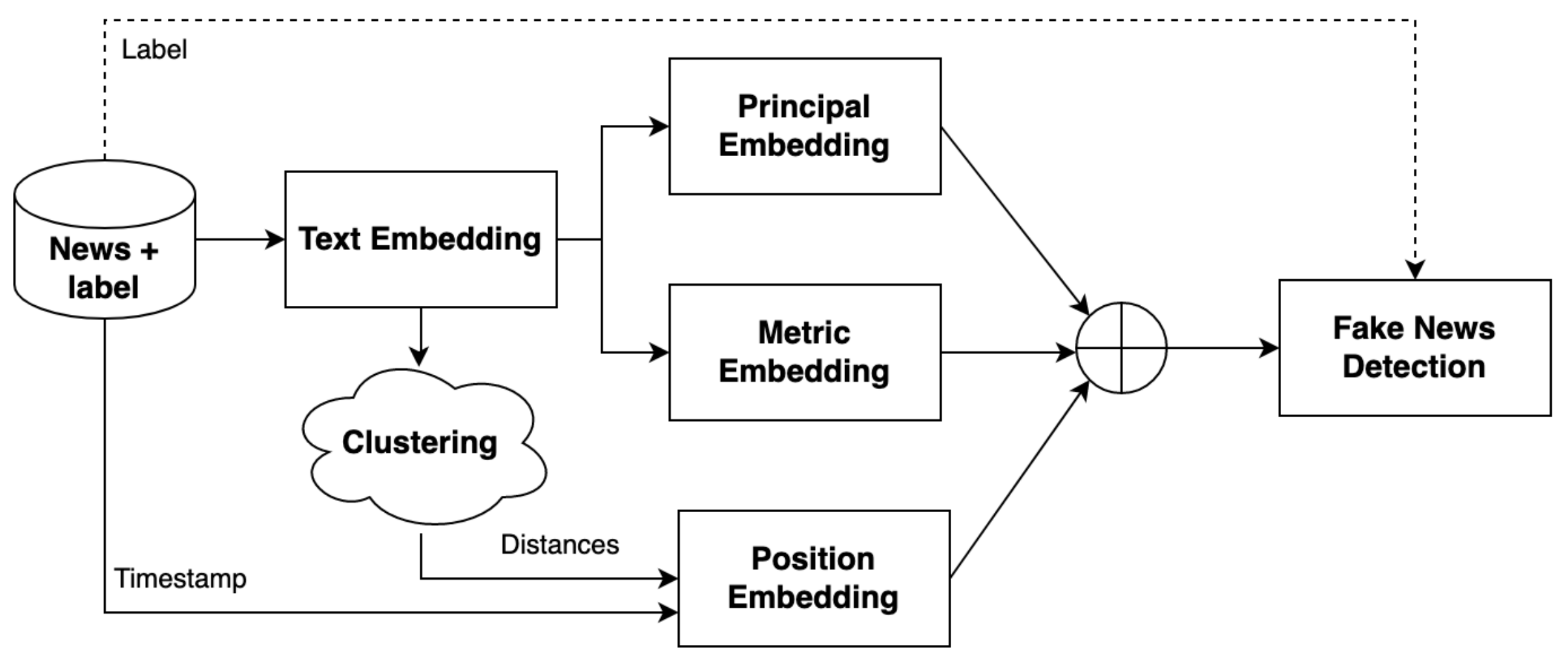}
\vspace{-6pt}
\caption{\FakeSwarm architecture.}
\label{fig:arch}
\end{figure}

\subsection{Text Embeddings}
Text embedding is a comprehensive task in NLP and includes various targeting embedding content at different levels, such as word embedding, sentence embedding, and document embedding. Word embedding is the basic fuel for all text embedding tasks because it transforms a word in the physical world into a word vector in semantic space. Higher-level embeddings are the aggregation and combination of word embeddings. \FakeSwarm considers the text embedding of news as a document embedding problem and utilizes word embeddings and pooling techniques to aggregate text embedding.

\subsubsection{Word Embedding}
Word embedding is a technique used in natural language processing to create numerical vectors that capture the semantic meaning of words. Word embeddings help to overcome the limitations of traditional bag-of-words models, which treat words as independent and unstructured units. With word embeddings, NLP models can more accurately understand the meaning of the text, leading to improved performance on a wide range of tasks.

\FakeSwarm selected Word2Vec \cite{mikolov2013efficient} as the word embedding algorithm because it performed better on the domain-specific corpus. The system takes the new body texts in \ISOT as the corpus. Each news is a long text with multiple sentences. The system first splits news into sentences through punctuation and then sets a hyperparameter to limit the maximal count of surrounding words for Word2Vec learning. Under the setting of embedding dimensionality, Word2Vec outputs a word embedding (a numerical vector format) for each word, comprehensive semantic information of all appearances of the words. 

\subsubsection{Text Embedding}
Nal \emph{et al.} \cite{kalchbrenner2014convolutional} achieved state-of-the-art results on several sentence classification tasks by using word embeddings as input and applied average pooling to obtain a fixed-size sentence embedding. Therefore, we also applied the \emph{average pooling} technique to generate the text embedding of news. Each dimension is the average of feature values in that dimension of all words in that news.

% 系统用 News embedding作为baseline方法，注意它只包含语义信息，而不包含swarm信息

\subsection{Principal Embedding}
Principal Component Analysis (PCA) is a technique used to reduce the dimensionality of high-dimensional data while retaining the most important information. The method works by finding a lower-dimensional representation of the data that captures as much of the variability in the data as possible.

The implementation of PCA involves calculating the eigenvectors and eigenvalues of the covariance matrix of the data. The eigenvectors represent the principal components of the data, while the eigenvalues represent the amount of variance captured by each principal component. The eigenvectors are sorted in descending order based on their corresponding eigenvalues, and the first few eigenvectors are selected to create the lower-dimensional representation of the data.

\FakeSwarm leverages PCA to generate new embeddings with a small count of dimensionality, which is called \emph{Principal Embedding}. Our system sets the default value of reduced dimensionality as a small number of 3. We infer fake news belonging to the same swarm would get closer in the space of \emph{Principal Embedding} because reducing dimensionality will reduce the global distances between every pair of data points and naturally reduces the distance of the data point inside each swarm. Evaluation results in sub section~\ref{subsec:eval_embedding} prove that \emph{Principal Embedding} effectively improves the detection accuracy.

\subsection{Metric Embedding}
%为什么要调整位置？期望让swarm之间更近，且分布地更加靠近
\FakeSwarm adopts contrastive learning to represent raw text embeddings, drawing inspiration mainly from our previous works: BotTriNet \cite{wu2023bottrinet} and FineEHR \cite{wu2023fineehr}. These works employed a metric-based approach to create sentence embeddings, enhancing the performance of downstream text classification tasks and surpassing the performance of the original sentence embeddings.

We consider contrastive learning could adjust the inner distances between fake news in the same swarm and define the represented embedding as the \emph{Metric Embedding}. Contrastive learning learns a new embedding from raw text embedding based on the optimizing object of reducing the distances between instances with the same label and also increasing the distances between instances having different labels. Naturally, fake news would be actively pushed together, the local density of each swarm would increase, and swarms would get far away from real news. 

\FakeSwarm uses a multilayer perceptron structure as the embedding network for representing the raw text embedding to the new Metric Embedding and also applies the contrastive loss to optimize the parameters of the perceptron. The contrastive loss is: \\
$Loss = Y * D^2 + (1-Y) * max(margin - D, 0)^2 $
\\
With $Y$, the label of two sampled instances (both are fake or real, label = 1; otherwise, label = 0), $D$ is the Euclidean distance between two metric embeddings. $margin$ is a hyper-parameter to adjust the distances between instances with different classes (fake or real).

\subsection{Position Embedding}
\label{design:selector}
% 我们打算用聚类算法，因为....
We made an assumption that news in a shared swarm will distribute close in the semantic space, and the distances between their text embedding are close too. However, mining features to measure the relationship between a data point, and its corresponding swarm is challenging due to several factors. First, the set of swarms (clusters) is varying and hard to decide. In other words, it is a traditional challenge in clustering problems. For example, the classic algorithm K-Means requires setting a hyperparameter of the number of clusters by hand. Second, correspondingly, the belonging swarm of a new is hard to decide. We consider simply bonding a news text embedding to the closest swarm (cluster) not enough due to the natural limitation of the clustering algorithm. 

To solve the first problem, \FakeSwarm chooses DBSCAN as the clustering algorithm. DBSCAN is particularly useful for datasets with irregular shapes and clusters of different densities, as it can identify clusters of any shape and size. The algorithm is also able to distinguish noise points that do not belong to any cluster. Also, it has the ability to automatically determine the number of clusters.

To solve the second challenge, \FakeSwarm not only involves the identification of belonging swarm as a positional feature but also extracts the distances from one news text embedding to all cluster centers respectively as extra positional features. The swarm identification and the distances make up the \emph{Positional Embedding}.

\subsection{Fake News Detection}
\FakeSwarm uses the concat of \emph{text embedding}, \emph{principal embedding}, \emph{metric embedding} and \emph{positional embedding} as a complete feature vector for fake news detection. We consider that concat embedding will improve the accuracy because it combines the semantic information and potential swarming characteristic for each piece of news. 

The system integrates various classifiers for fake news detection, e.g., Logistic Regression (LR), Multilayer Perceptron (MLP), Random Forest (RF), and Gradient Boosted Decision Tree (GBDT). The classifier uses the \emph{concat embeddings} as the feature vectors and uses the labels in the training data to fit a correlation function between the features and the label. The prediction process is similar, where the model uses the concat features of news to judge whether it is fake.

\section{Evaluation}
\label{sec:evaluation}

\subsection{Ground-truth and Metrics}

The concept of \emph{ground-truth} is vital in assessing the performance of predictive models using various metrics. In the \ISOT dataset, we divided all fake and real news articles into two portions using random splitting: a \emph{training set} (comprising 70 percent) and a \emph{testing set} (comprising 30 percent). Based on the predicted and actual labels of each sample, there are four possible outcomes: (i) True Positive (TP); (ii) False Positive (FP); (iii) True Negative (TN); (iv) False Negative (FN). We utilized two widely-accepted metrics, \emph{accuracy} and \emph{f1-score}, to assess the performance of \FakeSwarm in detecting fake news. \emph{Accuracy} is calculated as $\frac{TP+TN}{TP+FP+TN+FN}$ and represents the proportion of instances correctly classified, including both true and false instances in their respective categories. \emph{F1-score} is a combined measure of \emph{precision} and \emph{recall}, computed as $\frac{2 \cdot precision \cdot recall}{precision+recall}$. \emph{Precision}, defined as $\frac{TP}{TP+FP}$, measures the accuracy of identified positive instances. A higher f1-score indicates improved precision and recall.

\subsection{Effects of Swarming Characteristics}
\label{subsec:eval_embedding}

\begin{table*}[htbp]
\label{tab:eval_swarm}
  \centering
  \caption{Performance of Swarming Embeddings}
    \begin{tabular}{|c|c|c|c|c|c|c|c|c|}
    \hline
          & \multicolumn{2}{c|}{\textbf{GBDT}} & \multicolumn{2}{c|}{\textbf{RF}} & \multicolumn{2}{c|}{\textbf{LR}} & \multicolumn{2}{c|}{\textbf{MLP}} \\
    \hline
          & \textbf{Accuracy} & \textbf{F1score} & \textbf{Accuracy} & \textbf{F1score} & \textbf{Accuracy} & \textbf{F1score} & \textbf{Accuracy} & \textbf{F1score} \\
    \hline
    \textbf{Text} & 96.29\% & 96.45\% & 95.63\% & 95.84\% & 95.63\% & 95.84\% & 95.63\% & 95.84\% \\
    \hline
    \textbf{Prinicipal} & 93.27\% & 93.55\% & 92.97\% & 93.28\% & 93.00\% & 93.29\% & 92.91\% & 93.21\% \\
    \hline
    \textbf{Metric} & \textbf{96.85\%} & \textbf{96.99\%} & \textbf{97.33\%} & \textbf{97.45\%} & \textbf{96.60\%} & \textbf{96.75\%} & \textbf{96.55\%} & \textbf{96.69\%} \\
    \hline
    \textbf{Position} & 87.02\% & 87.55\% & 85.42\% & 86.40\% & 85.42\% & 86.40\% & 85.42\% & 86.40\% \\
    \hline
    \end{tabular}%
\end{table*}%

\begin{table*}[htbp]
\label{tab:eval_concat}
  \centering
  \caption{Performance of Concat Embeddings and FakeSwarm}
    \begin{tabular}{|c|c|c|c|c|c|c|c|c|}
    \hline
          & \multicolumn{2}{c|}{\textbf{GBDT}} & \multicolumn{2}{c|}{\textbf{RF}} & \multicolumn{2}{c|}{\textbf{LR}} & \multicolumn{2}{c|}{\textbf{MLP}} \\
    \hline
          & \textbf{Accuracy} & \textbf{F1score} & \textbf{Accuracy} & \textbf{F1score} & \textbf{Accuracy} & \textbf{F1score} & \textbf{Accuracy} & \textbf{F1score} \\
    \hline
    \textbf{Text+Prinicipal} & 96.31\% & 96.47\% & 97.25\% & 97.37\% & 95.96\% & 96.13\% & 96.47\% & 96.62\% \\
    \hline
    \textbf{Text+Metric} & 96.96\% & 97.09\% & 97.69\% & 97.80\% & \textbf{96.56\%} & \textbf{96.71\%} & 96.55\% & 96.68\% \\
    \hline
    \textbf{Text+Position} & 96.20\% & 96.36\% & 97.28\% & 97.40\% & 96.05\% & 96.22\% & 96.51\% & 96.65\% \\
    \hline
    \textbf{FakeSwarm(All)} & \textbf{96.98\%} & \textbf{97.11\%} & \textbf{97.84\%} & \textbf{97.94\%} & 96.54\% & 96.69\% & \textbf{96.70\%} & \textbf{96.84\%} \\
    \hline
    \textbf{Improvement} & \textcolor[rgb]{ 0,  .69,  .314}{\textbf{0.69\%}} & \textcolor[rgb]{ 0,  .69,  .314}{\textbf{0.67\%}} & \textcolor[rgb]{ 0,  .69,  .314}{\textbf{2.20\%}} & \textcolor[rgb]{ 0,  .69,  .314}{\textbf{2.11\%}} & \textcolor[rgb]{ 0,  .69,  .314}{\textbf{0.90\%}} & \textcolor[rgb]{ 0,  .69,  .314}{\textbf{0.85\%}} & \textcolor[rgb]{ 0,  .69,  .314}{\textbf{1.07\%}} & \textcolor[rgb]{ 0,  .69,  .314}{\textbf{1.01\%}} \\
    \hline
    \end{tabular}%
\end{table*}%

To address the unique characteristics of fake news networks, we devised three novel feature embeddings in addition to the baseline text embeddings: principal embeddings, metric embeddings, and positional embeddings. These new features aimed to enhance the recognition capabilities of our detection model. I conducted experiments to validate the effectiveness of each of these four embeddings individually in terms of accuracy and f1-score for fake news detection. For a comprehensive comparison, we applied four classic classification algorithms to test their performance on the testing dataset. Table~\ref{tab:eval_swarm} shows the result, where the performance of metric embedding is the best, even better than text embedding. Principal embedding also performs well. The result is reasonable because metric embedding and principal embedding are all representation of text embedding and retains semantic information. The performance of position embedding is not so good. But, think about it another way, it still achieved a very high accuracy of 87\% using pure swarming characteristic without direct text semantic information, and better than most classifications in \cite{ahmed2018detecting}.

We concatenated the text embeddings with various swarming embeddings and utilized a classifier for fake news detection to evaluate the performance of multimodal features. Incorporating multiple types of information has the potential to improve the detection capabilities of the model. The rationale behind this approach is that diverse embeddings can capture different aspects of fake news patterns, leading to a more comprehensive understanding of the underlying data. By combining text embeddings with swarming embeddings, we aim to extract richer features, thereby enhancing the classifier's ability to distinguish between real and fake news. Based on the experimental results shown in Table~\ref{tab:eval_concat}, we obtain several important conclusions. First, \FakeSwarm acquires the best performance on both f1score and accuracy by combining three kinds of swarming embeddings and text embeddings. Second, the concat of text embedding and arbitrary swarming embedding largely improved the performance compared with single text embedding. Third, the performance of position embedding was improved to the average level after combining with text embedding, proving that it is still effective for improving prediction accuracy even having no single-applying performance.

\subsection{Compare With Previous Works}
% Table generated by Excel2LaTeX from sheet 'related'
\begin{table}[htbp]
\label{tab:eval_previous}
  \centering
  \caption{Compare with Previous Works}
    \begin{tabular}{|c|c|c|}
    \hline
    Approaches & Accuracy & F1score \\
    \hline
    Ahmed et al. \cite{ahmed2018detecting} KNN & 83.00\% & / \\
    \hline
    Ahmed et al. \cite{ahmed2018detecting} DT & 89.00\% & / \\
    \hline
    Ahmed et al. \cite{ahmed2018detecting} LR & 89.00\% & / \\
    \hline
    Ahmed et al. \cite{ahmed2018detecting} SGD & 89.00\% & / \\
    \hline
    Ahmed et al. \cite{ahmed2018detecting} LSVM & 92.00\% & / \\
    \hline
    FakeSwarm Text & \textbf{95.63\%} & \textbf{95.84\%} \\
    \hline
    FakeSwarm Text + Swarming & \textbf{97.84\%} & \textbf{97.94\%} \\
    \hline
    \end{tabular}%
\end{table}%

We compare \FakeSwarm with the previous work \cite{ahmed2018detecting} that published \ISOT dataset. This research work applied the N-gram to generate tokens and used the TF-IDF metric to measure the frequency and importance of each gram. It also applied various classifiers to predict fake news and proved the Linear SVM achieved the best accuracy. The authors only show the accuracy score in the paper so we mainly compare it with our system using the accuracy score. 

Table~\ref{tab:eval_previous} shows the results. We conclude that text embedding has already been better than the best classifier in \cite{ahmed2018detecting}. Especially when combined with three swarming embeddings, \FakeSwarm achieved a very high accuracy with a very large improvement compared to the pure text embedding.

\subsection{Fake News Detection in Early Stage}

\begin{figure}[htp]
\centering
\label{fig:eval_month}
\begin{tabular}{cc}
\includegraphics[width=1.6in]{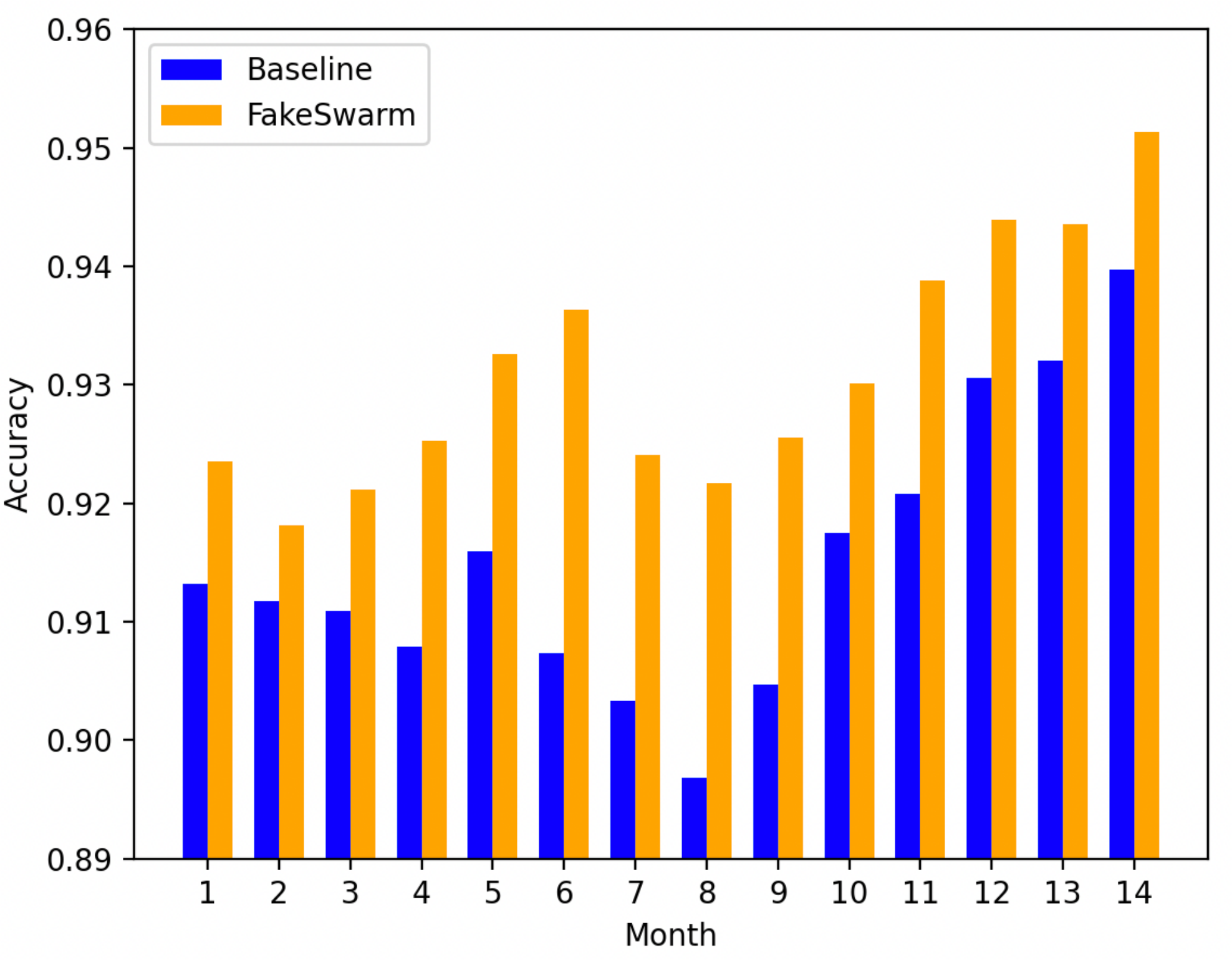} &
\includegraphics[width=1.6in]{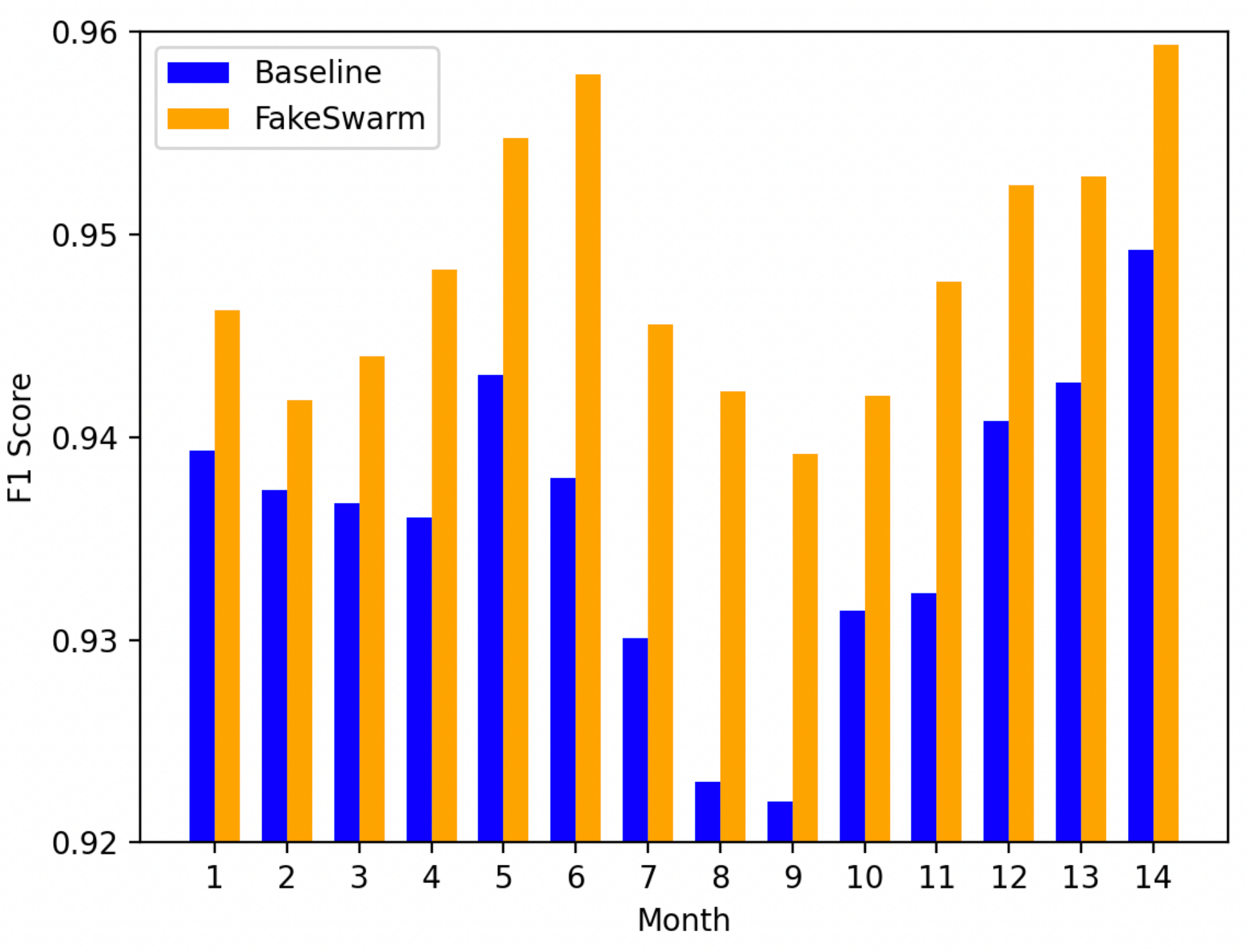}
\vspace{3pt}\\
\mbox{\small (a) Monthly Accuracy} &
\mbox{\small (b) Monthly F1score}
\end{tabular}
\vspace{-6pt}
\caption{Accuracy and F1score: \FakeSwarm and Baseline}
\label{fig:measure_day}
\vspace{-6pt}
\end{figure}

\begin{table*}[htbp]
\label{tab:eval_month_cnt}
  \centering
  \caption{The count of instances in monthly training and testing data set}
    \begin{tabular}{|c|c|c|c|c|c|c|c|c|c|c|c|c|c|c|}
    \hline
    \textbf{Month} & \textbf{1} & \textbf{2} & \textbf{3} & \textbf{4} & \textbf{5} & \textbf{6} & \textbf{7} & \textbf{8} & \textbf{9} & \textbf{10} & \textbf{11} & \textbf{12} & \textbf{13} & \textbf{14} \\
    \hline
    \textbf{Train Instances} & 1162  & 2424  & 3512  & 4637  & 5669  & 6646  & 7571  & 8601  & 9735  & 11048 & 12114 & 13538 & 14714 & 16114 \\
    \hline
    \textbf{Test Instances} & 4344  & 4056  & 3917  & 3668  & 3665  & 3862  & 4346  & 4392  & 4753  & 4582  & 4999  & 4568  & 4430  & 4050 \\
    \hline
    \end{tabular}%
\end{table*}%

Under the assumption that fake news is swarming appearing through the date time, we hope to simulate the fake news detection process in a real production environment. We construct streaming-format train and test data sets. In detail, we set a time parameter $month$, referring to the index of the month from the first detection date to the current detection date. We pick up 14 complete $month$ from \ISOT with stable news collection. The training set is all the news from the first month to the current month, and the testing set is the news published in the next month. \FakeSwarm uses the training set for text and swarming embedding generation and detects fake news appearing in the time window of the next month. 

Figure~\ref{fig:eval_month} shows the f1score and accuracy score varying through the increasing of $month$. \emph{Baseline} refers to using text embedding for fake detection, and \FakeSwarm refers to using text embedding and swarming embeddings for detection. The result shows that our system largely improves the detection accuracy at the early stage without enough training data (Table~\ref{tab:eval_month_cnt} shows the count of practical instances). And after the data starve stage, \FakeSwarm still has a stable improvement of accuracy compared to the baseline approach, which proves that the effectiveness of swarming embeddings is robust through time.

\section{Conclusions}
We introduced \FakeSwarm, a novel fake news identification system that utilizes swarming characteristics to enhance detection accuracy. By incorporating three types of swarm features, namely principal component analysis, metric representation, and position encoding, we demonstrated the effectiveness of considering swarming characteristics in fake news detection. 

Our evaluation on a public dataset revealed that combining all three types of swarm features achieved an impressive f1-score and accuracy of over 97\%, becoming the start-of-art detection system. We also developed an online learning pipeline to simulate the real production environment, we validated that our system still perform robust and accurate particularly in the early stages with limited training data.

In summary, \FakeSwarm introduces a fresh perspective and approach to fake news detection, emphasizing the significance of swarming characteristics in identifying and addressing the challenges posed by the proliferation of false information on social media platforms. We hope that future research will focus more on exploring swarming characteristics and investigating methods to further enhance the effectiveness of fake news detection.

\bibliographystyle{IEEEtran}
\bibliography{paper}

\end{document}